\documentclass[11pt,twoside]{article}

\usepackage{asp2004}
\usepackage{psfig}

\begin{document}
\title{Peculiar Red Nova V838 Mon:
Accretion and Interaction in a Wide Binary System
after Explosion of Its Companion}   

\author{Vitaly Goranskij and  Alla Zharova}
\affil{Sternberg Astronomical Institute, Moscow University, 119922, Russia}   

\author{Elena Barsukova, Sergei Fabrika, and Azamat Valeev}
\affil{Special Astrophysical Observatory, Russian Academy of Sciences,
Nizhnij Arkhyz, Karachai-Cherkesia, 369167, Russia}

\begin{abstract} 
We report the results of recent multicolor photometry and medium 
resolution spectroscopy of V838 Mon taken in 2007 -- 2008. In the eclipse-like
event in December 2006, the hot B3V type companion disappeared. The event 
accompanied by strengthening emission [FeII]/FeII lines in the spectra. We explain
this event as the formation of temporal short-lived accretion disc around
the hot companion. Later, in February 2007 the hot star reappeared in its
full brightness, but disappeared again for a long time since September 2007. 
This is the engulf of B3V companion by expanding remnant of 2002 outburst.
We assume that the thick accretion disc has formed around B type companion
which is moving now inside the envelope of the cool star. There is
some evidence of heating this disc and/or cool star envelope. We estimated the 
radius of expanding cool remnant in December, 2006 of about 150 A.U. or 
\hbox{30000 R$_\odot$.}
\end{abstract}

V838 Mon is a representative of a new enigmatic class of astrophysical
objects known as stars exploding into cool supergiants \citep{Munari+02},
or so-called red novae. Other members of this class are V4332 Sgr and V1006/7
in the galaxy M31. In such explosions, the matter of a few solar masses is
erupted into space, and therefore the ejected matherial does not reach 
optically thin state.
Before the outburst, V838 Mon was a system  of two B3V stars
\citep{Goranskij+07a}. The brightest star exploded in January 2002 and
reached M$_V$ = --10$^m$ in the light maximum having K -- M type spectrum.
The exploded star was young and chemically unevolved \citep{Kipper+07}, 
and it was brighter than its companion by factor of 1.36 $\pm$ 0.03 
\citep{Goranskij+07a}. Before the outburst, the exploded star had the 
reduced luminosity relative to normal B3V stars, its M$_V$ = +1$^m$.5.

The nature of such explosions is still unclear. Hypothesis of star collisions
and merging has still heavy grounds \citep{Socker+07}. In the beginning of
the XX century, this hypothesis was used to explain the explosions of
classical novae \citep{Flammarion+07}, and historically failed. 
In the case of V838 Mon, it needs a third star unseen in the spectral
energy distributions. \citet{Lynch+04} concluded that the exploded star was
G or F star evolved at least to the AGB stage, and perhaps beyond.
They report presence of Sr II and possibly other s-process elements.
They did not take into account binarity and based on the old unconfirmed
information on the progenitor. Moreover, exploded post-AGB stars like
FG Sge and V4334 Sgr are carbon-rich objects. The remnants of V838 Mon 
and V4332 Sgr are oxygen-rich stars \citep{Lynch+04,Barsukova+07}. 
On the contrary, \citet{Goranskij+07b} suppose that the explosion was 
caused by ignition of hydrogen in the centrum of a young massive star in 
the stage of gravitational contraction.

\begin{figure}
\plotfiddle{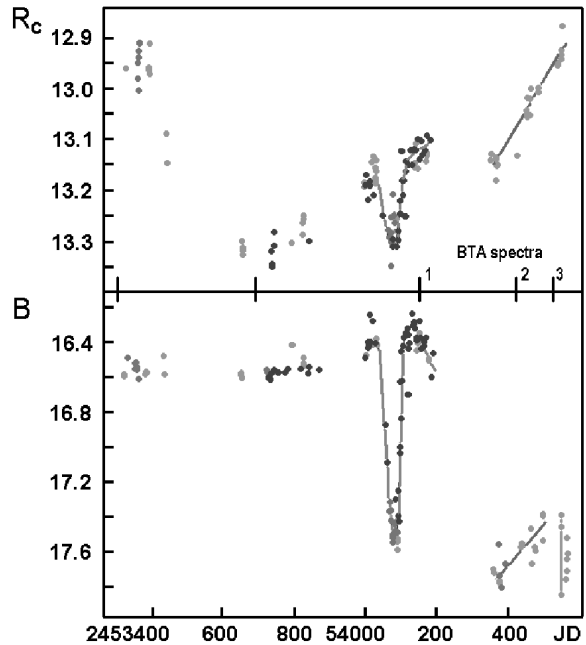}{7.8cm}{0}{43}{43}{-130}{-10}
\caption{Light curves of V838 Mon in 2004--2008 in the $B$ and $R$ bands.
Observations by \citet{Munari+07} are marked with dark points, and 
ours -- with grey points. Systematical differences between 
the observational sets were eliminated. Events shown: eclipse-like event
(curved lines); gradual brightness increase (straight lines); rapid decay 
in the $B$ band (vertical line). Dates of BTA spectroscopy are
pointed out by vertical line segments between the figure boxes}
\end{figure}

The expanding remnant of the outburst passes through an unique way of
evolution. \citet{Evans+03} using the 0.8 -- 2.5 $\mu$m spectra of V838 Mon
obtained on 2002 October 29 reported that the star became the coolest
supergiant -- ever observed, it might be the first known L type supergiant.
The $UBVRI$ photometry \hbox{by \citet{Goranskij+07b}} and \citet{Crause+05}
along with the spectroscopy by
\citet{Munari+05} confirmed that the remnant was so cool that its radiation
did not reach $UBV$ bands, the only hot companion radiation filled these bands.
A total of the remnant's flux was radiated in the red and infrared (IR) range.
Due to this fact, one can estimate correctly the contribution of the
exploded star in the common pre-outburst light of the binary in the $B$ and $V$
bands. The blackbody temperature of the remnant in the Autumn, 2002 was of
about 1200K, and increased to 1500K in 2004 \citep{Goranskij+07a}.
\citet{Pavlenko+07} estimated radius of the expanding remnant as large
as 6000 R$_\odot$ in November 2002, and this enormous value did not
contradict to the expansion velocity of 145 km/s of stellar photosphere
measured by \citet{Geballe+07} using high resolution IR spectra of first
overtone CO bands at 2.3 $\mu$m. Heating of such a large structure suggests
that the energy source may have a chemical nature -- burning of hydrogen,
carbon and metals in oxygen, what means in the simple sense a very large
conflagration in the body of the supergiant star.

\begin{figure}
\plotfiddle{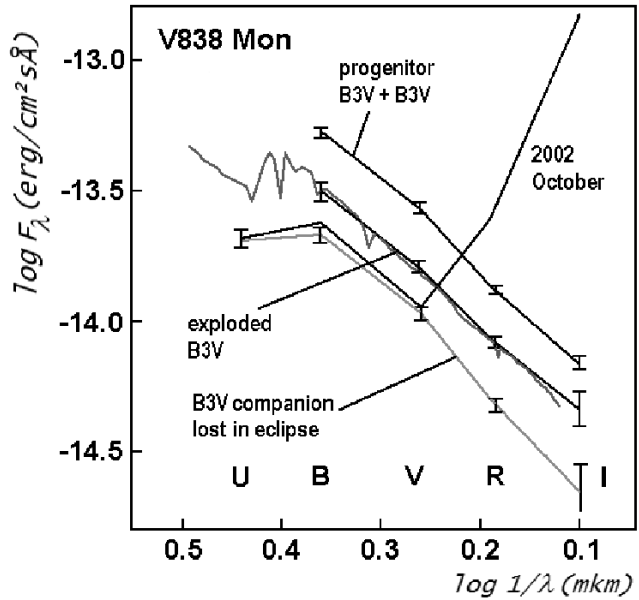}{7.5cm}{0}{40}{40}{-140}{-10}
\caption{Spectral energy distributions. From top to bottom: progenitor,
the binary system of two B3V stars; exploded brighter B3V star fitted by 
HD 29763 (B3V); binary combined light after outburst, in October 2002 
(with IR excess); B3V companion's SED disappeared in the eclipse-like event 
in December 2006 (grey line)}
\end{figure}      

At last, in the end of 2004 the signs of interaction of ejected material with 
the hot B3V companion became obvious, first, as rapid strengthening of iron
forbidden lines in the blue region of the spectrum \citep{Barsukova+06}.
Later, in 2006 very strong emission-line spectrum flared up with numerous
lines of [FeII]/FeII, Balmer, and [O I]. The flare of emissions was accompanied
with the temporal disappearance of the hot B3V type companion from the spectrum,
which looked like eclipse-like event in the short wavelength photometric bands
\citep{Goranskij06}.
The largest amplitude of light decay was of about 1$^m$.7 in $U$ band, and the
decay was imperceptible in $I_C$ band. The duration of event was of about
70 days with the light minimum at 2006 December 10, JD 2454080
\citep{Munari+07}. The early suggestions on the nature of this event
were the following: eclipse by a dust cloud, or by a blob thrown away in
the earlier outburst; grazing eclipse by the outer layers of the cool star
\citep{Munari+07}. \citet{Bond06} suggested that this was the engulf of
B3V companion by the expanding cool remnant. But Bond's suggestion contradicted
to reappearance of B3V star in February 2007.

In 2007 and 2008, we continued photometric and spectroscopic monitoring of
V838 Mon. The multicolor $UBVR_CI_C$ photometry was taken by V. Goranskij
and E. Barsukova with SAO 1-m Zeiss reflector. Additional photometry
in Johnson $BVR_JI_J$ bands was taken
by A. Zharova with 60-cm Zeiss reflector of
SAI Crimean Station. Our collection of photometry may be examined in detail with
Java-compatible browser in the Internet, the tables of observations 
are located there in the ASCII file (cf. \citet{Goranskij08}). The columns of 
the table contain the following data: JD hel.-- 2400000, $V, B, U, R, I$, 
remark to observation. Figure~1 shows $B$ and $R_C$ band light curves of
V838 Mon taken in the seasons of 2004 -- 2008. 

On 2007 September 18 in the beginning of the following 2007/2008 observational
season, we found V838 Mon in the deep minimum again, and this state
continued all this season. The new decay began probably just after rebrightening
in February 2007. The brightness in the new deep state was not constant.
It increased gradually in all the bands including long wavelength bands 
$R$ and $I$ (see e.g. Fig.~1 in $B$ and $R$).
Therefore, we do not treat this gradual increase as new reappearance of B3V type
companion. Rapid night-to-night variations in the short wavelength bands are
superimposed on the trend of such a rebrightening.
The biggest rapid decay was by 0$^m$.21 in $V$ and 0$^m$.43 in $B$ between the
nights of 27 and 28 March, 2008 (JD 2454553 -- 2454554) with no rapid change
in $R$ and $I$ bands. 

\begin{figure}
\plotfiddle{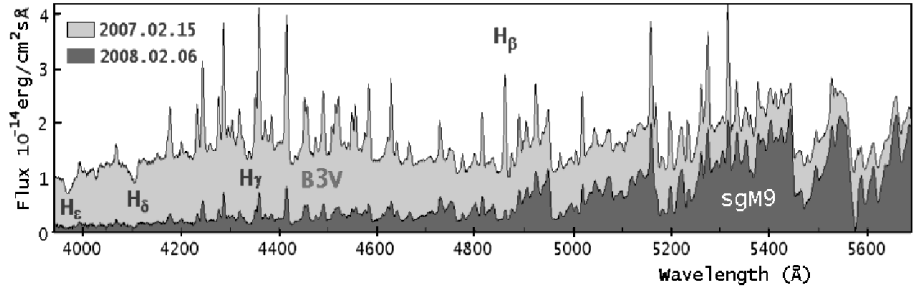}{3.8cm}{0}{42}{42}{-180}{-10}
\caption{Spectra of V838 Mon in blue and green range contain supergiant M9 and a
different amount of B3V companion light}
\end{figure}   

\begin{figure}
\plotfiddle{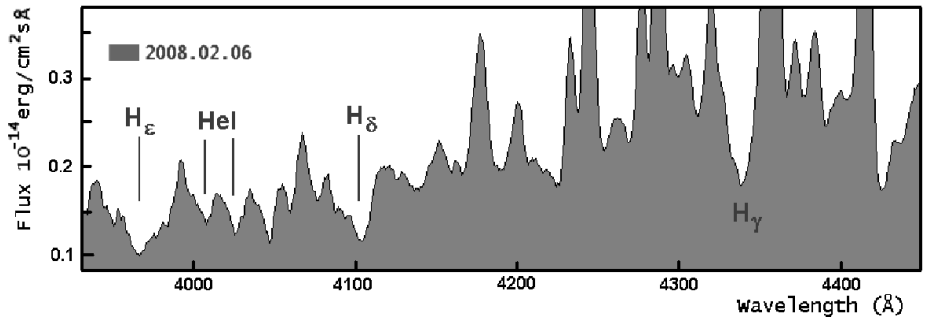}{3.8cm}{0}{43}{43}{-190}{-10}\caption
{The spectrum of reflected light of B3V companion}
\end{figure}

The spectrum of V838 Mon taken by \citet{Munari+07} on 2006 December 16 does not
show any traces of B3V type companion in the blue region. Therefore we can
measure the spectral energy distribution of B3V star in $UBVRI$ bands as the
light lost in the eclipse-like event. This estimate gives us a possibility
to measure energy distribution of the exploded star using archive pre-outburst
observations of progenitor binary in $BVRI$ bands. Earlier we obtained only
correct $B$ and $V$ magnitudes of exploded star using Autumn 2002
$BV$ photometry of
V838 Mon which did not contain the light of cool remnant in these bands
\citep{Goranskij+07a}. $R$ and $I$ magnitudes were contaminated by the light
of the cool remnant. In the Figure~2 we draw the spectral energy distribution
(SED) of progenitor B3V + B3V binary (upper black line), that one of common
light measured by us after the optical outburst, in Autumn 2002 (JD 2452530)
(black line with the red and IR excess), that one of the B3V companion light 
lost in eclipse-like event (grey line) and SED of exploded star calculated
as a flux difference of progenitor binary and flux of B3V companion lost in the 
eclipse (medium black line). The SED of exploded star is compared with the SED
of HD 29763 (B3V), and this fit is excellent. SEDs of V838 Mon are corrected
for interstellar absorption using color excess $E(B-V)$ = 0$^m$.77, and SED of
HD 29763 using $E(B-V)$ = 0$^m$.07 \citep{Goranskij+04}. Additionally, we see
good agreement of Autumn 2002 SED of B3V companion in UBV bands with the SED
of its lost light in eclipse-like event.

New spectral observations were taken with the Russian 6-m telescope BTA
and the SCORPIO spectrograph. The dates of spectra are marked in the Fig.~1
as vertical line segments, modern observations are pointed out by digits 1--3.
We have two spectra taken on 2007 February 15 (1) and 2008 February 6 (3)
covering the blue region of 3930--5700\AA\ with the resolution of 4.5 \AA,
and one spectrum taken on 2007 November 18 (2) covering all the optical range
between 3800 and 7570\AA\ with the resolution of 13 \AA. Spectral observations
taken with BTA were accompanied by $V$ band photometry.           

Unfortunately we did not take spectra during the eclipse-like
event in the middle
of December, 2006 due to bad weather. \citet{Munari+07} have obtained good
spectrum in the light minimum on 2006 December 26 using WHT (their Fig.~3).
These authors traced successively the changes in the spectrum before, in, and
after eclipse-like event (their Fig.~4). The observations showed that the
emission-line spectrum gradually strengthened independently on the event,
in spite of disappearance of B3V star, the main ionization and excitation source.
The maximum intensity of this spectrum was reached in February, 2007,
just after the recovery of B3V star. Our spectrum taken on 2007 February 15 is
shown in Figure~3 (light grey). In this spectrum, strong emissions of
[Fe II], Fe II and H$_\beta$ predominate. In the spectrum taken on 2008 February
6 (dark grey), these emission lines are essentialy depressed, and they
are about 10 times fainter than in February 2007. H$_\beta$ emission is
not seen or very faint. In this spectrum, very weak blue continuum of B3V type
companion is also present along with wide and intense Balmer absorptions and
weak HeI absorption lines (Figure~4). This weak blue spectrum seems not reddened
and may be reflected on the cloudy structure.

\begin{figure}
\plotfiddle{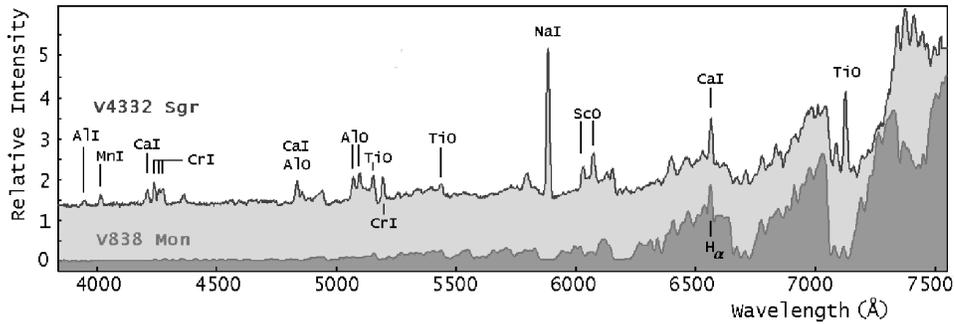}{4cm}{0}{44}{44}{-190}{-10}
\caption{Optical spectra of red novae V4332 Sgr and V838 Mon in 2007. The spectra are arbitrarily shifted along the vertical axis}
\end{figure}   

Otherwise, no traces of B3V star are seen in the spectrum with the lower
resolution of 13 \AA\ taken 3 month earlier, on 2007 Novebmer 18. There are weak
[FeII] emission lines and H$_\alpha$ emission.
The spectrum of cool companion looks like that one in 2005 January 15,
where the molecular bands were identified by \citet{Barsukova+07}. This spectrum
of V838 Mon is compared in Figure~5 with the spectrum of another red nova, 
V4332 Sgr taken with 6-m BTA telescope on 2007 June 18 with the same spectral 
range and resolution. In 2007, cool companions of these novae
reached equal temperature, and their spectra looked very similar. We classified
them as M9 type supergiants. This similarity was the result of evolution from 
the opposite directions: the remnant of V838 Mon warmed up, but the remnant 
of V4332 Sgr cooled down during its two-year decay by 2$^m$ in $V$ band. 
Note, that spectral emission-line components in these novae have principally 
different nature. This is the radiation of hot ionized gas in V838 Mon, and
the radiation of cool rarefied atomic and molecular gas in V4332 Sgr.

We assume that the December 2006 light decay was due to formation of highly
inclined temporal accretion disc around B3V companion. The rim of this disc
might overlap the light of the hot star only for a terrestrial
observer, but did not block the ionizing radiation in the both directions
above and below the disc plane. The primordial disc might be formed from
the ejected
matter moving ahead of expanding photosphere of the cool remnant and
infalling into
the Roche lobe of the B3V star with acceleration, and it was rapidly heated
by the hot companion, and dissipated.

Photometry shows that the ejected matter of 2002 outburst riched B3V companion 
for 1750 days and then formed the accretion disc around it. Using the photosphere
expansion velocity of 145 km/s measured by \citet{Geballe+07} we can estimate
the radius of the cool remnant to be equal of 30000 R$_\odot$ in December 2006.
The distance between companions in this wide system before the explosion of one
of them was of about 150 A.U. Assuming the circular orbit, we have raw count
of $\sim$500 years for the orbital period of such a binary, and the orbital
velocity of about 10 km/s.  

In the following season of 2007/2008 B3V type companion disappeared for a
long time. This phenomemon seems to be real engulf of the B3V companion by the
expanding cool star. Now, B3V star is moving under the remnant's photosphere.
The rapid light variability in the short wavelength bands, reflected light
of the blue companion, and presence of forbidden FeII lines in the spectrum
suggest that B3V star is seen through cloudy structure of the external shell
with an inner void. Such a model was calculated by \citet{Lynch+07}.
Presence of forbidden lines in the spectrum confirms that the small part of the
ionizing radiation of B3V star leaks out through cloudy shell.

Another consequence of engulf is gradual increase of brightness in all the
photometric bands which began in the late 2006, before the eclipse-like event.
The rise of brightness continues during 2007/2008 season.
The increasing excess has red energy distribution. We assume that new thick
accretion disc is accumulating around B3V star which moves inside the envelope
of cool remnant. The excess of radiation may be explained by heating of this
disc and surrounding matter by B3V type star. It is hard to predict further
development of the V838 Mon system. The red novae are binary
systems, and this fact is well established. The evolution of red novae remnants
may pass through individual ways depending on the structure of the systems
and the dimensions of their orbits.\\

\acknowledgements
V.G. and S.F. thank the chair person of the Conference, Elena P. Pavlenko and the
Organizing Committee for invitation to the Conference and good organization.
This research was supported by the Russian Foundation of Basic Research by
Grants No. 03-02-16133, 05-02-19710-JF, 06-02-16865, 07-02-00630.

\end{document}